\documentclass[9pt,shortpaper,twoside,web]{ieeecolor}
\usepackage{generic}
\usepackage{amssymb}
\usepackage{amsmath}
\usepackage{color}
\newtheorem{definition}{Definition}
\newtheorem{assumption}{Assumption}
\newtheorem{theorem}{Theorem}
\newtheorem{remark}{Remark}
\newtheorem{lemma}{Lemma}
\newtheorem{proposition}{Proposition}
\newtheorem{claim}{Claim}
\usepackage{bm}
\usepackage{graphicx}
\usepackage{epstopdf}
\usepackage{cite}
\usepackage{algorithm,algpseudocode,float}
\usepackage{lipsum}

\makeatletter
\newenvironment{breakablealgorithm}
  {% \begin{breakablealgorithm}
   \begin{center}
     \refstepcounter{algorithm}% New algorithm
     \hrule height.8pt depth0pt \kern2pt% \@fs@pre for \@fs@ruled
     \renewcommand{\caption}[2][\relax]{% Make a new \caption
       {\raggedright\textbf{\ALG@name~\thealgorithm} ##2\par}%
       \ifx\relax##1\relax % #1 is \relax
         \addcontentsline{loa}{algorithm}{\protect\numberline{\thealgorithm}##2}%
       \else % #1 is not \relax
         \addcontentsline{loa}{algorithm}{\protect\numberline{\thealgorithm}##1}%
       \fi
       \kern2pt\hrule\kern2pt
     }
  }{% \end{breakablealgorithm}
     \kern2pt\hrule\relax% \@fs@post for \@fs@ruled
   \end{center}
  }
\makeatother
\def\BibTeX{{\rm B\kern-.05em{\sc i\kern-.025em b}\kern-.08em
    T\kern-.1667em\lower.7ex\hbox{E}\kern-.125emX}}
\begin{document}

\title{\LARGE \bf
Linear Convergent Distributed Nash Equilibrium Seeking with Compression}
\author{Xiaomeng Chen$^{1}$, \thanks{$^{1}$X. Chen, Y. Wu and L. Shi are with the Department of Electronic and Computer Engineering, Hong Kong University of Science and Technology, Clear Water Bay, Kowloon, Hong Kong (email: \{xchendu,eewuyc,eesling\}@ust.hk).} Yuchi Wu$^{1}$,  Xinlei Yi$^{3}$, \thanks{$^{3}$X. Yi is  with the Division of Decision and Control
Systems, School of Electrical Engineering and Computer Science, KTH
Royal Institute of Technology, and also affiliated with the Digital Futures,
Stockholm 10044, Sweden (e-mail: xinleiy@kth.se).} Minyi Huang$^{4}$,\thanks{$^{4}$M. Huang is  with the School of Mathematics and Statistics, Carleton University, Ottawa, ON K1S 5B6, Canada (e-mail:
mhuang@math.carleton.ca).} and Ling Shi$^{1}$
}

\maketitle

\begin{abstract}
Information compression techniques are majorly employed to address the concern  of reducing communication cost over peer-to-peer links.  In this paper, we investigate distributed Nash equilibrium (NE) seeking problems in a class of non-cooperative games over {\color{black}directed graphs} with information compression.  To improve  communication efficiency, a compressed distributed NE seeking (C-DNES) algorithm  is  proposed to obtain a NE for games, where the differences between decision vectors and their estimates are compressed. The proposed algorithm is compatible with a general class of compression operators, including  both unbiased and biased compressors.  {\color{black}Moreover, our approach  only requires the adjacency matrix of the directed graph to be row-stochastic, in contrast to past works that relied on balancedness or specific global network parameters.} It is shown that C-DNES not only inherits the advantages of  conventional  distributed NE algorithms, achieving linear convergence rate for games with restricted strongly monotone mappings, but also saves communication costs in terms of transmitted bits. Finally, {\color{black}numerical simulations  illustrate  the advantages of C-DNES in saving communication cost by an order of magnitude under different compressors.}
\end{abstract}

%%%%%%%%%%%%%%%%%%%%%%%%%%%%%%%%%%%%%%%%%%%%%%%%%%%%%%%%%%%%%%%%%%%%%%%%%%%%%%%%
\section{INTRODUCTION}
 Game theory has been studied extensively  on account of its significant role in analyzing the  interactions among rational decision-makers. Engineering applications of game-theoretic methods include congestion control in traffic networks \cite{ma2011decentralized}, charging or discharging of electric vechicles \cite{grammatico2017dynamic} and demand-side management in smart grid \cite{saad2012game}.
As an important issue, Nash equilibrium (NE) seeking has attracted ever-increasing attention with the emergence of multi-agent networks. Conceptually, NE is a proposed solution in multiplayer noncooperative games, where a number of selfish players aim to minimize their own cost functions  by making decisions according to others' actions. 

A large body of work on NE seeking  has been reported in the recent literature (see e.g.\cite{yu2017distributed,belgioioso2018projected,shamma2005dynamic} and references therein). Most  of them assumed that each player can access  the actions of all other players. This global knowledge assumption requires a central coordinator to broadcast the information to the network, which is sometimes impractical \cite{ghaderi2014opinion,bimpikis2019cournot}. Hence, distributed NE seeking algorithms
have become a subject of great interest recently, where a distributed information sharing protocol is adopted to exchange local messages among players. For example, Ye et al. \cite{ye2017distributed} designed distributed NE seeking method based on consensus protocols, where each player estimates and eventually reconstructs the action profile  of all players. For constrained aggregate games, a subgradient-based distributed algorithm was proposed  by Lou et al. \cite{lou2015Nash} for NE seeking over time-varying netowrks. In  discrete time cases,  early works \cite{koshal2016distributed,salehisadaghiani2016distributed}   only studied algorithms with vanishing step-sizes, resulting in slow convergence. Recently, fixed step-size algorithms were designed by Tatarenko et al. \cite{tatarenko2020geometric} based on  monotonicity property, guaranteeing  linear convergence for NE seeking. Bianchi and Grammatico \cite{bianchi2020Nash} designed a fixed-step gradient algorithm
to seek a NE over  directed graphs with the knowledge of Perron-Frobenius (PF) eigenvector of the adjacency matrix. 
 
\begin{figure}[htp]
\centering
 \includegraphics[width=0.5\textwidth]{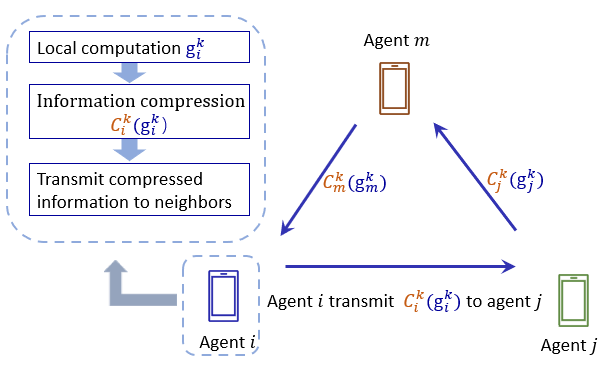}
  \caption{Distributed framework with compressed information at each time iteration $k$ in a multi-agent system of $3$ agents.}
      \label{compressed}
\end{figure}

 All the approaches mentioned above assume  unlimited communication bandwidth. However, in practical systems such as underwater vehicles and low-cost unmanned aerial vehicles, the communication capacity and bandwidth are often limited by the environments. 
 As the dimension of data increases, the burden of information exchange between players will cause  communication bottleneck, which deteriorates  the efficiency of the algorithm. Thus, in order to fulfill the requirement of limited bandwidth, a compressor is necessary for each agent to send compressed information that is encoded with fewer number of bits. {\color{black}Before transmitting  local messages, each agent firstly compresses the information then transmits the compressed one to their neighbors, as shown Fig. \ref{compressed}.} Recently, various  information compression methods, such as quantization and sparsification, have been adopted in distributed optimization in centralized networks \cite{seide20141,alistarh2017qsgd,bernstein2018signsgd,karimireddy2019error,mishchenko2019distributed,liu2020double}. For decentralized optimization, with certain compression error feedback techniques, some works achieved linear convergence rate for optimization algorithms with compressed information \cite{liao2021compressed,kovalev2021linearly,liu2020linear,zhang2021innovation}.

 Though the distributed NE seeking problem may be viewed as an extension of distributed optimization problems, the existing information compression methods cannot be directly applied to distributed NE seeking   problems due to the fact that the objective function of each player  in noncooperative games relies on the actions of all players. Owing to the more complex information exchange among players, such as estimation of  joint action profile, more difficulties are faced when designing the adaptive compression approach. Thus,  few works studied distributed NE seeking problem with compressed communication. 
  Nekouei et al. \cite{nekouei2016performance}   investigated the impact of quantized communication on the behavior of NE seeking algorithms over fully connected communication graphs, which is usually  an impractical assumption in distributed networks.  A continuous-time distributed NE seeking algorithm with finite communication bandwidth is proposed by Chen et al. \cite{chen2022distributed}, where a specific quantizer is used.   Recently, logarithmic and uniform quantizations are adopted by Ye et al. \cite{ye2022distributed} to develop a quantized NE seeking strategy in continuous-time systems. However, the proposed method in \cite{ye2022distributed} only converge to a neighborhood of NE point. 

\indent All the above motivates us to further develop a discrete-time distributed NE seeking algorithm with information compressed by general compressors before transmission. To the best of our knowledge, this paper is the first to propose a  communication-efficient discrete-time distributed algorithm over directed graphs, which can converge linearly to an NE for noncooperative games  under a general class of compressors. 
	The main contributions of this paper are summarized as follows:

\begin{enumerate}
{\color{black}
\item In designing a compressed distributed NE seeking algorithm (C-DNES), we consider a general class of compressors instead of specific compression schemes in \cite{chen2022distributed,ye2022distributed}. The assumption in this paper encompasses both biased and unbiased compressors, as well as norm-sign compressors and the composition of quantization and sparsification.

\item In contrast to earlier studies \cite{koshal2016distributed,salehisadaghiani2016distributed,tatarenko2020geometric,bianchi2020Nash,chen2022distributed} that relied on balancedness or awareness of specific global  network parameters (PF eigenvector), our method only necessitates the  row-stochasticity assumption for the adjacency matrix of the communication graph. This mild  assumption in our proposed C-DNES framework allows for the accommodation of both undirected and directed topology graphs, which is advantageous for practical implementation.
\item  We prove that there exists a unique NE  for  games with  restricted strongly monotone mappings (unlike the strongly monotone mappings in \cite{chen2022distributed,ye2022distributed}) and C-DNES is guaranteed to linearly converge to the unique NE. This relatively moderate assumption of game mapping expands  applicability of  C-DNES to a more extensive range of games.
\item  Through simulation examples, we illustrate that the proposed  C-DNES algorithm is applicable to various compressors and has   convergence performance comparable to those of state-of-the-art algorithms with accurate communication. Meanwhile,  we show that C-DNES  can decrease the transmitted bits by an order of magnitude. }

\end{enumerate}

 \textit{Notations:}  In this paper, $\bm{1}\in\mathbb{R}^n$  represents the column vector with each entry given by $1$. We denote $\bm{e}_i \in\mathbb{R}^n$ as the $i$-th unit vector which takes zero except for the $i$-th entry that equals to $1$. $\mathbb{R}_{++}$ denotes the set of all positive real numbers. The spectral radius of matrix $\mathbf{A}$ is denoted by $\rho(\mathbf{A})$. The smallest nonzero eigenvalue of a positive semidefinite matrix $ \mathbf{M} \succeq  \mathbf{0}$ is denoted as $\tilde \lambda_{\min}(\mathbf{M})$. Given a vector $\mathbf{x}$, we denote its $i$-th
element by $[\bm{x}]_{i}$. Given a matrix $\mathbf{A}$, we denote its
element in the $i$-th row and $j$-th column by $[\bm{A}]_{ij}$. Let sgn($\cdot$) and $|\cdot|$ be the element-wise sign function and absolute value function, respectively. We denote  $||\mathbf{x}||$ and $||\mathbf{X}||_{\text{F}}$ respectively as $l_2$ norm of vector $\mathbf{x}$ and the Frobenius norm of matrix $\mathbf{X}$. 
 We denote  by $\mathbf{A} \odot \mathbf{B}$, the Hadamard product of two matrices, $\mathbf{A}$ and $\mathbf{B}$. In addition, $\bm{x} \preceq \bm{y}$ is denoted as component-wise inequality between vectors $\bm{x}$ and $\bm{y}$.  For a matrix $\mathbf{A} \in \mathbb{R}^{n\times n}$, denote $\text{diag}(\mathbf{A})$ as its diagonal vector, i.e., $\text{diag}(\mathbf{A})=[[\bm{A}]_{11},\ldots, [\bm{A}]_{nn}]^\top$. For a vector $\mathbf{a} \in \mathbb{R}^{n}$, denote $\text{Diag}(\mathbf{a})$ as the diagonal matrix with the vector $\mathbf{a}$ on its diagonal. A matrix $\mathbf{A} \in \mathbb{R}^{n\times n}$ is consensual if it has equal row vectors. 

\section{PRELIMINARIES AND PROBLEM STATEMENT}
\subsection{Network Model}
We consider a group of agents  communicating with each other over {\color{black}a directed graph} $G\triangleq (\mathcal{V}, \mathcal{E})$, where $\mathcal{V}=\{1,2,3\ldots, n\}$ denotes the agent set and $\mathcal{E}\subset \mathcal{V} \times \mathcal{V}$ denotes the edge set, respectively. A communication link from agent $j$ to agent $i$ is denoted by $(j,i)\in \mathcal{ E}$, indicating that agent $j$ can send messages to agent $i$. The set of neighbors of agent $i$ is denoted as $\mathcal{N}_i= \{j\in \mathcal{V} | (j,i) \in  \mathcal{ E}\}$. The adjacency matrix of the graph is denoted as $W=[w_{ij}]_{n\times n}$ with $w_{ij}>0$ if $(i,j)\in \mathcal{ E}$ or $i=j$, and $w_{ij}=0$ otherwise. {\color{black}The graph $G$ is called strongly connected if there exists at least
one directed path from any agent $i$ to any agent $j$ in the directed graph with $i\neq j$. 
\begin{assumption}\label{asp3}
	The directed graph $G\triangleq (\mathcal{V}, \mathcal{E})$ is strongly connected. Moreover, the adjacency matrix $W$ associated with $G$ is row-stochastic, i.e., $\sum_{l=1}^n w_{il}=1, \forall i \in \mathcal{V}$.
		\end{assumption}}

\subsection{Compressors}
{\color{black}A stochastic compressor $\mathcal{C}(\mathbf{x}, \xi):  \mathbb{R}^d \times \mathcal{Z}  \rightarrow \mathbb{R}^d$ is a mapping that convert  a $\mathbb{R}^d$-valued signal $\mathbf{x}$ to a compressed one, where  $\mathbf{\xi}$ is a  random variable with range $\mathcal{Z}$. Note that the realizations $\xi$ of the compressor $\mathcal{C}(\cdot)$ are independent  among different agents and time steps. In other words, given an sequence $\{\mathbf{x}_i^{k}\}$ for each agent $i \in \mathcal{V}$, and any time iteration $k\geq 0$, the randomness $\xi_i^k$ with each usage of compressor $\mathcal{C}_i^k(\mathbf{x}_i^k,\mathbf{\xi}_i^k)$ are i.i.d, where $\mathcal{C}_i^k(\cdot)$ denotes the compressor used by agent $i$ at iteration $k$.  For deterministic compressors, which can be treated as special cases of the random ones, the random variable $\xi$ is not included, thus the notation reduces to $\mathcal{C}(\mathbf{x})$. For stochastic compressors, the notation $\mathbb{E}_{\xi}$ is used to  denote  the expectation over the  inherent randomness  in the stochastic compressor \cite{singh2022sparq,singh2021squarm,yi2022communication,koloskova2019decentralized1}. For deterministic compressors, the expected value operator will return the deterministic argument. Hereafter, we drop $\xi$ and write $\mathcal{C}(\mathbf{x})$ for notational simplicity. }

Next, we introduce a  general assumption on the  compression operators which is considered in our paper.
{\color{black}
\begin{assumption}\label{c3}
	For any agent $i \in \mathcal{V}$ and any iteration $k\ge0$, 
 the compression operator $\mathcal{C}_i^k$ satisfies
\begin{equation}
	\mathbb{E}_{\xi}[||\mathcal{C}_i^k(\mathbf{x})-\mathbf{x}||^2|\mathbf{x}]\leq C||\mathbf{x}||^2,\qquad \forall \mathbf{x}\in\mathbb{R}^d,
\end{equation}
where $C\geq 0$ and the r-scaling of $\mathcal{C}_i^k$ satisfies
\begin{equation}
	\mathbb{E}_{\xi}[||\mathcal{C}_i^{k}(\mathbf{x})/r-\mathbf{x}||^2|\mathbf{x}]\leq (1-\delta)||\mathbf{x}||^2,\qquad \forall \mathbf{x}\in\mathbb{R}^d, 
\end{equation}
where   $\delta \in (0,1]$ and $r>0$.
\end{assumption}

\begin{remark}Assumption \ref{c3} requires the mean square of the \textit{relative} compression error to be bounded. Note that if  
$\mathcal{C}$ is unbiased, i.e., $\mathbb{E}_{\xi}[\mathcal{C}(\mathbf{x})|\mathbf{x}]=\mathbf{x}$, Assumption \ref{c3} degenerates to the condition of unbiased compressors in  \cite{alistarh2017qsgd},\cite{mishchenko2019distributed},\cite{liu2020double},\cite{liu2020linear},\cite{wen2017terngrad}. Moreover, if $0\le C<1$, compressors satisfying Assumption \ref{c3} is equivalent to the class of biased but contractive compressors which are widely adopted in practice \cite{koloskova2019decentralized,stich2020communication,beznosikov2020biased}. In a word, Assumption \ref{c3} is a general assumption of unbiased and biased compressors.
\end{remark}}

Some commonly used compressors satisfying Assumption \ref{c3} are given below.

\textbf{1) Unbiased compressor: Stochastic quantization}:

The $b-$bit $q-$norm quantization compressor is defined as follows:
\begin{equation}\label{q1}
	\mathcal{C}_q(\mathbf{x})=\frac{||\mathbf{x}||_q}{2^{b-1}} sgn(\mathbf{x})\odot\left\lfloor \frac{2^{b-1}|\mathbf{x}|}{||\mathbf{x}||_q}+\mathbf{u}\right\rfloor,
\end{equation}
where  $\mathbf{u}$ is a random vector uniformly sampled from $[0,1]^d$, which is independent of $\mathbf{x}$.  For $q=\infty$, the compressor satisfies Assumption \ref{c3} with $C=d/4^{b}$. Owing to the fact that only the norm $||\mathbf{x}||_q$, sgn($\mathbf{x}$) and integers in the bracket  need to be transmitted, this compressor is widely adopted in compressed distributed learning  \cite{mishchenko2019distributed,liu2020double,wen2017terngrad} with $l_\infty$ norm quantization, while $l_{2}$ norm quantization is adopted in \cite{alistarh2017qsgd,kovalev2021linearly,koloskova2019decentralized}.

\textbf{2) Biased but contractive compressor: Top-$\mathbf{k}$ sparsification}:

The largest $k$ coordinates in magnitude is selected among the vector $\mathbf{x}$, i.e.,
\begin{equation}\label{topk}
	\mathcal{C}_{top}(\mathbf{x})=\mathbf{x}\odot \mathbf{e},
\end{equation} 
where the elements of $\mathbf{x}$ is reordered as $[\mathbf{x}]_1\geq[\mathbf{x}]_2\geq \cdots [\mathbf{x}]_d$ and $\mathbf{e}$ satisfies $[\mathbf{e}]_l=1$ for $l\leq k$ and $[\mathbf{e}]_l=0$ for $l> k$. This compressor meets Assumption \ref{c3} with $C=1-k/d$.

\textbf{3) General compressor: Norm-sign compressor}:

{\color{black}\begin{equation}\label{normsign}
	\mathcal{C}( \mathbf{x})=||\mathbf{x}||_q sgn(\mathbf{x}), \quad q\geq2.
\end{equation}  
\begin{claim}\label{claim}
	For the compressor defined in \eqref{normsign}, we have the following the compression constants satisfying Assumption \ref{c3}.
\begin{equation*}
	\begin{matrix}
		C=d-1,&r=d,&\delta =\frac{1}{d}. 	\end{matrix}
\end{equation*}
\end{claim}
\begin{proof}
   See Appendix	\ref{cp}.
\end{proof}

\subsection{Problem Statement}
Consider a  noncooperative game  in a multi-agent system of $n$ agents, where 
each agent has an unconstrained action set $\Omega_i=\mathbb{R}$. Without loss of generality, we assume that  each agent's decision variable is a scalar. Let $J_i$ denote  the cost function of the agent $i$. Then, the game is denoted as $\Gamma(n, \{J_{i}\}, \{\Omega_i=\mathbb{R}\})$.   The goal of each agent $i\in \mathcal{V}$ is to minimize its objective function $J_i(x_i,x_{-i})$, which depends on both the local variable $x_i$ and the decision variables of the other agents $x_{-i}$. We then make the following assumptions with respect to game $\Gamma$. 
\begin{assumption}\label{cvx}
For all $i \in\mathcal{V}$, the  cost function $J_i(x_i,x_{-i})$ is strongly convex and continuously differentiable in $x_i$ for each fixed $x_{-i}$. 
\end{assumption}
%\begin{assumption}\label{cvx}
%The game $\Gamma(n, \{J_{i}\}, \{\Omega_i\})$ is convex, i.e., for all $i \in\mathcal{V}$, the set $\Omega_i$ is nonempty, compact and convex
%, the  cost function $J_i(x_i,x_{-i})$ is convex and continuously differentiable in $x_i$ for each fixed $x_{-i}$. 
%\end{assumption}

%Then the game mapping is defined as 
%\begin{equation}
%	\mathbf{f(x)}\triangleq [\nabla_1 J_1(x_1,x_{-1}),\ldots, \nabla_n J_n(x_n,x_{-n})]^\top,
%\end{equation} 
%where $\nabla_i J_i(x_i,x_{-i})=\frac{\partial J_i(x_i,x_{-i})}{\partial x_i},\forall i \in\mathcal{V}$. 

\begin{definition}
	The mapping $\mathbf{F}: \mathbb{R}^n \rightarrow \mathbb{R}^n$, referred to be the game mapping of $\Gamma(n, \{J_{i}\}, \{\Omega_i \})$ is denoted as
\begin{equation}
	\mathbf{F(x)}\triangleq [\nabla_1 J_1(x_1,x_{-1}),\ldots, \nabla_n J_n(x_n,x_{-n})]^\top,
\end{equation} 
where $\nabla_i J_i(x_i,x_{-i})=\frac{\partial J_i(x_i,x_{-i})}{\partial x_i},\forall i \in\mathcal{V}$. 
\end{definition}

{\color{black}\begin{assumption}\label{asp1}
	The game mapping $\mathbf{F(x)}$ is restricted strongly monotone  to any NE $\mathbf{x}^\star $ with constant $\mu_r>0$, i.e.,
	$$\langle \mathbf{F(x)}-\mathbf{F(\mathbf{x}^\star)},\mathbf{x}-\mathbf{\mathbf{x}^\star}  \rangle \geq \mu_r ||\mathbf{x-\mathbf{x}^\star}||^2, \forall \mathbf{x} \in\mathbb{R}^n.$$
		 
\end{assumption}

\begin{remark}
In this paper, we relax the strongly monotone mapping condition in \cite{yu2017distributed,ye2017distributed,tatarenko2020geometric}  to a restricted one as Assumption \ref{asp1}. Thus, a wider variety of games are taken into consideration. \end{remark}}

%{\color{black}\begin{remark}
%Assumption \ref{asp1} is widely used for general Nash games \cite{yu2017distributed,ye2017distributed,tatarenko2020geometric} aiming at establishing convergence of algorithms to compute a NE point. In case of potential games \cite{monderer1996potential} which is a vital class of Nash games, the Jacobian of the game mapping $\mathbf{F}: \mathbb{R}^n \rightarrow \mathbb{R}^n$ is symmetric \cite{facchinei2003finite}. Hence, the game mapping corresponds to  gradient of a single function, which is called potential function of the game. In terms of potential games, the monotonicity condition of gradient mapping is equivalent to the convexity condition of the potential function. Neyman \cite{neyman1997correlated} illustrated that if the potential function $\mathbf{P}$ of the game is convex, then the set of minimizers of $\mathbf{P}$ equals the set of NE of the game. In this case, the problem of learning a Nash equilibrium reduces to a single-objective optimization. The strictly convex condition for potential function  is assumed  in recent works of distributed NE seeking in potential games \cite{tatarenko2018learning,huang2020distributed}. However, in our paper, we do not restrict ourselves to the potential games, i.e., the game mapping $\mathbf{F}$ may have non-symmetric Jacobian. 
%\end{remark}
%}

\begin{assumption}\label{asp2}
	Each function $\nabla_i J_i(x_i,x_{-i}) $ is Lipschitz continuous in $x_i$  for every fixed $x_{-i}$,  i.e.,  $\exists L_i \geq 0$, we have $\forall x_i, y_i,$	$$|\nabla_i J_i(x_i,x_{-i})-\nabla_i J_i(y_i,x_{-i}) |\leq L_i |x_i-y_i|.$$
	
	Moreover, each function $\nabla_i J_i(x_i,x_{-i}) $ is Lipschitz continuous in $x_{-i}$ for every fixed $x_{i}$,  i.e.,  $\exists L_{-i} \geq 0$, we have $\forall x_{-i}, y_{-i},$	$$|\nabla_i J_i(x_i,x_{-i})-\nabla_i J_i(x_i,y_{-i}) |\leq L_{-i} |x_{-i}-y_{-i}|.$$

\end{assumption}

%Assumption \ref{asp1} and \ref{asp2} are required for linear convergence of algorithms  computing an equilibrium point in variational inequalities and games \cite{tatarenko2020geometric},\cite{belgioioso2019distributed}.  
The concept of NE is given below. 
\begin{definition} 
	A vector $\mathbf{x}^\star=[x_1^\star,x_2^\star,\ldots,x_n^\star]^\top $ is a NE if for any $i \in \mathcal{V}$ and $x_i \in\mathbb{R}$,
	\begin{equation}\label{ne1}
		J_i(x_i^\star, x_{-i}^\star)\leq J_i(x_i, x_{-i}^\star).
	\end{equation}
\end{definition}

In this paper, we are interested in distributed seeking of a NE in a game $\Gamma(n, \{J_{i}\}, \{\Omega_i =\mathbb{R}\})$ where Assumption \ref{asp3},\ref{cvx},\ref{asp1} and \ref{asp2} hold. Note that  a NE  can be alternatively characterized by using the first-order optimality conditions \cite{facchinei2003finite}. To be specific,  $\mathbf{x}^\star \in \mathbb{R}^n$ is a NE if and only if for all $x_i$, we have 
\begin{equation}\label{ne}
	\langle \nabla J_i(x_i^\star, x_{-i}^\star), x_i-x_i^\star \rangle \ge 0. 
\end{equation}
%%%%%%%%%%%%%%%%%%%%%%%%%%%%%
\section{COMPRESSED DISTRIBUTED NASH EQUILIBRIUM SEEKING}
\subsection{Nash Equilibrium Seeking in Distributed Settings }
To cope with incomplete information, we assume that each agent maintains a local variable 
\begin{equation}
	\mathbf{x}_{(i)}=[\tilde x_{(i)1},\ldots,\tilde x_{(i)i-1},x_i,\tilde x_{(i)i+1},\ldots,\tilde x_{(i)n}]^\top \in \mathbb{R}^n,
\end{equation}
 which is its estimation of the joint action profile $\mathbf{x}=[x_1,x_2,
 \ldots,x_n]^\top$, where $\tilde x_{(i)j}\in\mathbb{R}$ denotes agent $i$'s estimate of $x_j$ and $\tilde x_{(i)i}=x_i\in \mathbb{R}$.  
 
% Moreover, we denote the estimation of other players' action as
% 
%\begin{equation}
%	\mathbf{x}_{(i)}=[\tilde x_{(i)1},\ldots,\tilde x_{(i)i-1},x_i,\tilde x_{(i)i+1},\ldots,\tilde x_{(i)n}]^\top \in \mathbb{R}^n,
%\end{equation}
 Denote estimation matrix, the compact form of action-profile estimates from all agents, as $$\mathbf{X}=[\mathbf{x}_{(1)},\mathbf{x}_{(2 )},\ldots,\mathbf{x}_{(n)}]^\top\in \mathbb{R}^{n \times n},$$
 where the $i$th row is  the estimation vector $\mathbf{x}_{(i)}, i\in\mathcal{V}$.  At $k-$th iteration, action-profile estimates $\mathbf{X}$ are denoted by $\mathbf{X}^{k}$.

 Moreover, for any given action-profile estimates, we define a diagonal matrix 
 $$\mathbf{\tilde F}(\mathbf{X})\triangleq \text{Diag}(\nabla_1 J_1(\mathbf{x}_{(1)}),\ldots, \nabla_n J_n(\mathbf{x}_{(n)}))\in \mathbb{R}^{n \times n}.$$
 
 {\color{black}Next, we present the following proposition showing a equivalent condition for NE in the game $\Gamma(n, \{J_{i}\}, \{\Omega_i\})$.
 
 \begin{proposition} (\hspace{-0.001cm}\cite{tatarenko2020geometric1}) \label{proposition} Consider the game $\Gamma(n, \{J_{i}\}, \{\Omega_i\})$ satisfies Assumption \ref{asp3} and  \ref{cvx}. Then the following statements are equivalent
 \begin{enumerate}
 	\item The vector $\mathbf{x}^\star=[x_1^\star,\ldots, x_n^\star]^\top$ is a NE in $\Gamma(n, \{J_{i}\}, \{\Omega_i\})$.   
 	\item There exists an estimation matrix $\mathbf{X}^\star $ with the diagonal vector $\mathbf{x}^\star $ and the corresponding diagonal matrix $\mathbf{\tilde F}(\mathbf{X}^\star)$ such that for any $\mathbf{X} $ the following holds
 	 \begin{equation}\label{vi}
 		 \langle (I-W)\mathbf{X}^\star+\eta\mathbf{\tilde F}(\mathbf{X}^\star), \mathbf{X}^\star-\mathbf{X} \rangle \ge 0,
 			\end{equation}
 	 where $\eta >0$ is an arbitrary constant. 
 \end{enumerate}
 	
 \end{proposition} 
 
 Proposition \ref{proposition} shows that any solution matrix $\mathbf{X}^\star$ of the variational inequality \eqref{vi} is consensual and its diagonal vector  is a NE in $\Gamma(n, \{J_{i}\}, \{\Omega_i\})$. Furthermore, from the variational inequality \eqref{vi}, we can define the matrix mapping $\mathbf{F}_a: \mathbb{R}^{n \times n} \rightarrow \mathbb{R}^{n \times n}$ as
 \begin{equation}\label{augmap}
 	\mathbf{F}_a(\mathbf{X})=(I-W)\mathbf{X}+\eta\mathbf{\tilde F}(\mathbf{X}),   \end{equation}
 which is referred to as the \textit{augmented mapping}  \cite{tatarenko2020geometric1} of game $\Gamma(n, \{J_{i}\}, \{\Omega_i\})$. }

\subsection{Distributed Nash Equilibrium Seeking with Information Compression }
With the above analysis for NE seeking in distributed networks, we are now ready to introduce our proposed compressed algorithm, where  compressors $\mathcal{C}(\cdot)$ satisfying Assumption \ref{c3} are adopted  to propose a communication-efficient  algorithm to find a NE of the game $\Gamma$ in a fully distributed manner. The agents aim to asymptotically reconstruct the true values of the actions of the other agents, based on the compressed data received from their neighbors. 
 
 The detailed procedures are presented   in Algorithm \ref{alg1} and {\color{black}the notations used throughout Algorithm \ref{alg1} are summarized in Table \ref{table}. }

 \begin{breakablealgorithm}
\caption{A Compressed Distributed Nash Equilibrium Seeking (C-DNES) Algorithm}
\label{alg1}
\hspace*{-0.1in} {\bf Input:} %
stopping time $K$, step-size $\eta$, consensus step-size $\gamma$, scaling parameters $\alpha>0$, and initial values $\mathbf{x}_{(i)}^0, \mathbf{h}_{i}^0$\\
\hspace*{-2.75in} {\bf Output: $\mathbf{x}_{(i)}^K$}
\begin{algorithmic}[1]
\For {each agent $i\in\mathcal{V}$}
\State $\mathbf{h}_{i,w}^0=\sum\limits_{j\in \mathcal{N}_i} w_{ij}\mathbf{h}_{j}^0$
\EndFor
\For {$k=0,1,2,\ldots, K-1$} locally at each agent $i\in\mathcal{V}$

\State  
$\mathbf{q}_i^k=\mathcal{C}_i^k(\mathbf{x}_{(i)}^k-\mathbf{h}_{i}^k)$   .
\State $\mathbf{\hat x}_{(i)}^k=\mathbf{h}_{i}^k+\mathbf{q}_i^k$
\State $\mathbf{h}_{i}^{k+1}=(1-\alpha)\mathbf{h}_{i}^k+\alpha \mathbf{\hat x}_{(i)}^k$
\State Send $\mathbf{q}_i^k$ to agent $l \in \mathcal{N}_i$ and receive $\mathbf{q}_j^k$ from agent $j \in \mathcal{N}_i$ 
\State $\mathbf{\hat x}_{(i),w}^k =\mathbf{h}_{i,w}^k+\sum_{j=1}^n w_{ij}\mathbf{q}_j^k$
\State $\mathbf{h}_{i,w}^{k+1}=(1-\alpha)\mathbf{h}_{i,w}^k+\alpha\mathbf{\hat x}_{(i),w}^k$
\State $\mathbf{x}_{(i)}^{k+1}=\mathbf{x}_{(i)}^{k}-\gamma(\mathbf{\hat x}_{(i)}^k-\mathbf{\hat x}_{(i),w}^k)-\gamma\eta\nabla_i J_i(\mathbf{x}_{(i)}^{k})\mathbf{e}_i$
%\State Update \vspace*{-7mm}
%
%\begin{equation*}	\begin{aligned}
%		&\mathbf{\hat x}_{(i),w}^k =\mathbf{h}_{i,w}^k+\sum_{j=1}^n w_{ij}\mathbf{q}_j^k\\
%		&\mathbf{h}_{i,w}^{k+1}=(1-\alpha)\mathbf{h}_{i,w}^k+\alpha\mathbf{\hat x}_{(i),w}^k\\
%		&\mathbf{x}_{(i)}^{k+1}=\mathbf{x}_{(i)}^{k}-\gamma(\mathbf{\hat x}_{(i)}^k-\mathbf{\hat x}_{(i),w}^k)-\eta\nabla_i J_i(\mathbf{x}_{(i)}^{k})\mathbf{e}_i
%	\end{aligned}
%\end{equation*}
\EndFor
 \end{algorithmic}
\end{breakablealgorithm}
 \begin{table}[H]
 \caption{Descriptions of notations in  Algorithm \ref{alg1}}
 \label{table}
\centering
{\color{black}\begin{tabular}{cc}
\hline
Symbol & Description \\ \hline
\qquad$\mathbf{x}_{(i)}$             \quad         &  Local estimation of the joint action profile    \quad    \\

\qquad$\mathbf{\hat x}_{(i)}$   \quad                        &      {\color{black} Estimated  version of $\mathbf{x}_{(i)}$ }     \\
\qquad$\mathbf{\hat x}_{(i),w}$ \quad                         &      Weighted average of $\mathbf{\hat x}_{(i)}$     \\
\qquad$\mathbf{h}_{i}$   \quad                        &      {\color{black} Reference point of $\mathbf{x}_{(i)}$ }      \\

\qquad$\mathbf{h}_{i,w}$  \quad                         &      Weighted average of $\mathbf{h}_{i}$       \\
\qquad$\mathbf{q}_{i}$   \quad                       &   Transmitted compressed value of $\mathbf{x}_{(i)}-\mathbf{h}_{i}$     \\ 
\hline
\end{tabular}}

\end{table}

% The function \mathcal{C} denotes the compressor that compresses the variables for each agent at every iteration. 
{\color{black}\begin{remark}
 In C-DNES, instead of compressing the local variable $\mathbf{x}_{(i)}$, we maintain an auxiliary variable $\mathbf{h}_{i}^k$, acting as \textit{a reference point} of $\mathbf{x}_{(i)}$, and compress the difference $\mathbf{x}_{(i)}^k-\mathbf{h}_{i}^k$.  As $\mathbf{h}_{i}^k$ approaches $\mathbf{x}_{i}^k$, by Assumption \ref{c3}, the variance of compression error will tends  to $0$. After receiving the compressed value $\mathbf{q}_i^k=\mathcal{C}(\mathbf{x}_{(i)}^k-\mathbf{h}_{i}^k)$, each agent $i$ obtains \textit{an estimator} of $\mathbf{x}_{(i)}$, $\mathbf{\hat x}_{(i)}$ by assembling from $\mathbf{h}_{i}^k$ and the received value.  Then $\mathbf{h}_{i}^{k+1}$ is obtained as the weighted average of its previous value $\mathbf{h}_{i}^k$  and $\mathbf{\hat x}_{(i)}^k$ with mixing weight $\alpha$, indicating that $\mathbf{h}_{i}^k$ is tracking the motions of $\mathbf{x}_{i}^k$. The update procedure of $\mathbf{h}_{i}^k$ is motivated from  DIANA \cite{mishchenko2019distributed} and LEAD \cite{liu2020linear}, which  controls the compression error, particularly for a relatively large
constant $C$ in Assumption  \ref{c3}. Moreover, the variable $\mathbf{h}_{i,w}$ is a \textit{weighted averaged} version of $\mathbf{h}_{i}$, which can be regarded as a backup copy for the neighboring information. The introduction of this auxiliary variable eliminates the need to store all the neighbors' variable $\mathbf{h}_{j}$.
\end{remark}}

% In Line 5, the difference between $\mathbf{x}_{(i)}^k$ and the auxiliary variable $\mathbf{h}_{i}^k$ is compressed and then added back to $\mathbf{h}_{i}^k$ in Line 6 in order to  obtain $\mathbf{\hat x}_{(i)}^k$. Then $\mathbf{h}_{i}^{k+1}$ is obtained as the weighted average of its previous value $\mathbf{h}_{i}^k$  and $\mathbf{\hat x}_{(i)}^k$ with mixing weight $\alpha$. 

 Denote the compact form of stochastic approximation of action estimates  as $$\mathbf{\hat X}=[\mathbf{\hat x}_{1},\mathbf{\hat x}_{2 },\ldots,\mathbf{\hat x}_{n}]^\top\in \mathbb{R}^{n \times n},$$
 where the $i$th row is  the approximation vector $\mathbf{\hat x}_{i}, i\in\mathcal{V}$. Auxiliary variables  in compact form at $k-$th iteration are denoted as $\mathbf{\hat X}^{k}, \mathbf{H}^{k},\mathbf{H}_w^{k}, \mathbf{Q}^k$ and $ \mathbf{\hat X}^{k}_w$ , respectively.

 Algorithm \ref{alg1} can be written in compact form as follows:
 \begin{subequations}
 \begin{align}\label{ala}
&\mathbf{Q}^k=\mathcal{C}^k(\mathbf{X}^{k}-\mathbf{H}^{k}),\\ \label{alb} 
&\mathbf{\hat X}^{k}=\mathbf{H}^{k}+\mathbf{Q}^k,\\ 
&\mathbf{\hat X}^{k}_w=\mathbf{H}_w^{k}+W\mathbf{Q}^k,\\\label{ald}
 &\mathbf{H}^{k+1}=(1-\alpha)\mathbf{H}^{k}+\alpha\mathbf{\hat X}^{k},\\
  &\mathbf{H}^{k+1}_w=(1-\alpha)\mathbf{H}^{k}_w+\alpha\mathbf{\hat X}^{k}_w,\\
&\mathbf{X}^{k+1}=\mathbf{X}^{k}-\gamma(\mathbf{\hat X}^{k}-\mathbf{\hat X_w}^{k})-\gamma\eta \mathbf{\tilde F}(\mathbf{X}^{k}),\label{alf} 
 \end{align}
\end{subequations}
 where $\mathbf{X}^{0}$ and $\mathbf{H}^{0}$ are arbitrary chosen.  

Note that after initialization $\mathbf{H}^{0}_w=W\mathbf{H}^{0}$, we have 
\begin{equation}
\begin{aligned}
	\mathbf{H}^{1}_w&=(1-\alpha)\mathbf{H}^{0}_w+\alpha(\mathbf{H}_w^{0}+W\mathbf{Q}^0)\\
	&=W[(1-\alpha)\mathbf{H}^{0}+\alpha\mathbf{\hat X}^{0}]=W\mathbf{H}^{1}.
	\end{aligned}
\end{equation}

%$\mathbf{\hat X}^{0}_w=\mathbf{H}_w^{0}+W\mathbf{Q}^0=W(\mathbf{H}^{0}+\mathbf{Q}^0)=W\mathbf{\hat X}^{0}$  and  $\mathbf{H}^{1}_w=(1-\alpha)\mathbf{H}^{0}_w+\alpha\mathbf{\hat X}^{0}_w=W[(1-\alpha)\mathbf{H}^{0}+\alpha\mathbf{\hat X}^{0}]=W\mathbf{H}^{1}$. 
Hence, by induction, we obtain $\mathbf{H}^{k}_w=W\mathbf{H}^{k}$ and $\mathbf{\hat X}^{k}_w=W\mathbf{\hat X}^{k}$  for all $k$.  Then, the state variable update in \eqref{alf} becomes
\begin{equation}\label{algx}
	\begin{aligned}
		&\mathbf{X}^{k+1}=\mathbf{X}^{k}-\gamma(\mathbf{\hat X}^{k}-W\mathbf{\hat X}^{k})-\gamma\eta \mathbf{\tilde F}(\mathbf{X}^{k})\\
		&=(1-\gamma)\mathbf{X}^{k}+\gamma W\mathbf{\hat X}^{k}-\gamma\eta \mathbf{\tilde F}(\mathbf{X}^{k})\\
		&=\mathbf{X}^{k}-\gamma(I-W)(\mathbf{X}^{k}-\mathbf{E}^{k})-\eta\gamma \mathbf{\tilde F}(\mathbf{X}^{k})\\
		&={\color{black}\mathbf{X}^{k}-\gamma \mathbf{F}_a(\mathbf{X}^{k})+\gamma(I-W)\mathbf{ E}^{k},}\\
	\end{aligned}
\end{equation}where $\mathbf{ E}^{k}=\mathbf{X}^{k}-\mathbf{\hat X}^{k}$ denotes the compression error for the decision variable. {\color{black}The consensus step-size $\gamma$ ensures the algorithmic convergence, which is proved theoretically in Section \ref{cov}. }
\begin{remark}
	It is worth noting that the above equation implies that C-DNES performs an implicit error compensation mechanism that alleviates the impact of compression error. The $(I-W)\mathbf{ E}^{k}$ term in \eqref{algx} shows that each agent $i$ transmits compression error to its neighbors and compensates this error locally by adding $\bm{e}_i^k$, where $\bm{e}_i^k \in \mathbb{R}^n$ is the $i-$th row of $\mathbf{ E}^{k}$. 
\end{remark}
%	If the algorithm performs no communication compression, i.e., $\mathbf{\hat X}^{k}=\mathbf{X}^{k}$, the decision update rule reduces to
%	\begin{equation}
%	\begin{aligned}
%		\mathbf{X}^{k+1}&=\mathbf{X}^{k}-\gamma(I-W)\mathbf{ X}^{k}-\gamma \eta \mathbf{\tilde F}(\mathbf{X}^{k})\\		&=\mathbf{X}^{k}-\gamma \mathbf{F}_a(\mathbf{X}^{k}),\\
%	\end{aligned}
%\end{equation}
% recovering the distributed NE seeking algorithm in \cite{tatarenko2020geometric1}. 
%	

 \section{CONVERGENCE ANALYSIS}\label{cov}
 In this section, we analyze the convergence performance of C-DNES. 
 The main idea of our strategy is to bound $\mathbb{E}_{\xi}[||\mathbf{X}^{k+1}-\mathbf{X}^\star||^2_{\text{F}}]$ and $\mathbb{E}_{\xi}[||\mathbf{X}^{k+1}-\mathbf{H}^{k+1}||^2_{\text{F}}]$ on the basis of the linear combinations of their previous values. By establishing a linear system of inequalities, we can derive the convergence result.

Let $\mathcal{F}^{k}$ be the $\sigma-$algebra generated by $\{\mathbf{ E}^{0},\mathbf{ E}^{1}, \ldots, \mathbf{E}^{k-1}\}$, and denote $\mathbb{E}_{\xi}[\cdot\mid \mathcal{F}^{k}]$ as the conditional expectation  given $\mathcal{F}^{k}$. {\color{black}The following flow graph illustrates the relation between iterative variables in  Algorithm \ref{alg1}, where the solid arrows shows the dynamics of the algorithm.
  
\begin{figure}[htp]
\centering
 \includegraphics[width=0.5\textwidth]{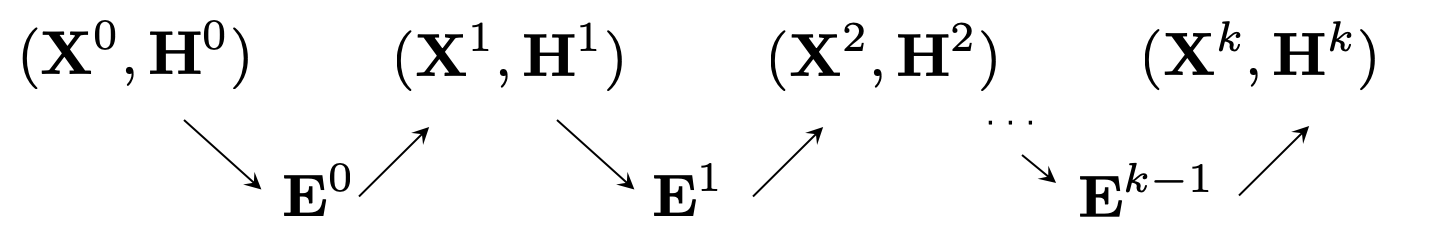}
  \caption{Flow chart of variables during iterations.}
      \label{fg1}
\end{figure}

Since the inherent randomness of the compressor is not correlated across the iteration steps, we can obtain
%\begin{equation}\label{sequence}
%\begin{aligned}
%&\mathbb{E}_{\xi}[||\mathbf{ X}^{k}-\mathbf{H}^k-\mathcal{C}^k(\mathbf{X}^{k}-\mathbf{H}^{k})||^2_{\text{F}}\mid \mathcal{F}^{k-1}]	\\
%&\leq C||\mathbf{X}^{k}-\mathbf{H}^{k}||^2_{\text{F}}. \\
%\end{aligned}
%\end{equation} }

\begin{equation}\label{sequence}
\begin{aligned}
&\mathbb{E}_{\xi}[||\mathbf{ X}^{k}-\mathbf{H}^k-\mathcal{C}^k(\mathbf{X}^{k}-\mathbf{H}^{k})||^2_{\text{F}}\mid (\mathbf{ X}^{k},\mathbf{H}^{k})]	\\
=&\mathbb{E}_{\xi}[||\mathbf{ X}^{k}-\mathbf{H}^k-\mathcal{C}^k(\mathbf{X}^{k}-\mathbf{H}^{k})||^2_{\text{F}}\mid \{(\mathbf{ X}^{j},\mathbf{H}^{j})\}_{j=0}^{k}]	\\
=&\mathbb{E}_{\xi}[||\mathbf{ X}^{k}-\mathbf{H}^k-\mathcal{C}^k(\mathbf{X}^{k}-\mathbf{H}^{k})||^2_{\text{F}}\mid \mathcal{F}^{k-1}]\\
\leq &C||\mathbf{X}^{k}-\mathbf{H}^{k}||^2_{\text{F}}. \\
\end{aligned}
\end{equation}

\subsection{Technical lemmas}
  We first prepare a few  supporting lemmas for further convergence analysis.

\begin{lemma}\label{lmm}
	The variables $\mathbf{ X}^{k}, \mathbf{H}^{k}$ are measurable  with respect to $\mathcal{F}^{k}$. Furthermore, we have
	\begin{equation}
		\mathbb{E}_{\xi}[||\mathbf{E}^{k}||^2_{\text{F}}\mid \mathcal{F}^{k-1}]\leq C||\mathbf{X}^{k}-\mathbf{H}^{k}||^2_{\text{F}}.
	\end{equation}
\end{lemma}
\begin{proof}
	By expanding \eqref{ala}-\eqref{alf} recursively, $\mathbf{ X}^{k}$ and $ \mathbf{H}^{k}$ can be  represented by linear combinations of $\mathbf{ X}^{0}, \mathbf{H}^{0}$ and random variables $\{\mathbf{E}^j\}_{j=0}^{k-1}$, i.e, $\mathbf{ X}^{k}, \mathbf{H}^{k}$ are measurable  respect to $\mathcal{F}^{k}$. Moreover, from Equation \eqref{alb} and \eqref{sequence}, we have  
\begin{equation}\label{comp}
\begin{aligned}
\mathbb{E}_{\xi}&[||\mathbf{E}^{k}||^2_{\text{F}}\mid \mathcal{F}^{k-1}]=\mathbb{E}_{\xi}[||\mathbf{ X}^{k}-\mathbf{\hat X}^k||^2_{\text{F}}\mid \mathcal{F}^{k-1}]\\
&=\mathbb{E}_{\xi}[||\mathbf{ X}^{k}-\mathbf{H}^k-\mathcal{C}^k(\mathbf{X}^{k}-\mathbf{H}^{k})||^2_{\text{F}}\mid \mathcal{F}^{k-1}]	\\
&\leq C||\mathbf{X}^{k}-\mathbf{H}^{k}||^2_{\text{F}}. \\
\end{aligned} 
\end{equation} 
\end{proof}
  
	\begin{lemma}\label{lmeq}
		(Lemma 7 in \cite{liao2021compressed}) For any $\mathbf{U,V} \in \mathbb{R}^{n\times n}$, the following inequality is satisfied:
		\begin{equation}
			||\mathbf{U}+\mathbf{V}||^2\leq \tau'||\mathbf{U}||^2+\frac{\tau'}{\tau'-1}||\mathbf{V}||^2,
		\end{equation}
		where $\tau'>1$. Moreover, for any $\mathbf{U}_1, \mathbf{U}_2, \mathbf{U}_3 \in \mathbb{R}^{n\times n}$, we have $||\mathbf{U}_1+\mathbf{U}_2+ \mathbf{U}_3||^2\leq \tau_1 ||\mathbf{U}_1||^2+\frac{2\tau_1}{\tau_1-1}(||\mathbf{U}_2||^2+||\mathbf{U}_3||^2)$ and $||\mathbf{U}_1+\mathbf{U}_2+ \mathbf{U}_3||^2\leq 3 ||\mathbf{U}_1||^2+3||\mathbf{U}_2||^2+3||\mathbf{U}_3||^2$. 
	\end{lemma}

{\color{black}\begin{lemma} \label{lml}(Lemma 1 in \cite{tatarenko2020geometric1})
Given Assumption \ref{asp2}, the augmented mapping $\mathbf{F}_a$	of game $\Gamma(n, \{J_{i}\}, \{\Omega_i\})$ is Lipschitz continuous  with  $L_F=\eta L_m+||I-W||_{\text{F}}$, where $L_m=\max_i\{\sqrt{L_i^2+L_{-i}^2}\}$. 
\end{lemma}
\begin{lemma}\label{lmmu} (Lemma 3 in \cite{tatarenko2020geometric1})
Given Assumption \ref{asp3}, \ref{asp1} and \ref{asp2}, the augmented mapping $\mathbf{F}_a$	of game $\Gamma(n, \{J_{i}\}, \{\Omega_i\})$ is restricted strongly monotone to any NE matrix $\mathbf{X}^\star=\mathbf{1}(\mathbf{x}^\star)^\top$ with the constant $\mu_F=\min\{b_1,b_2\}>0$, where $b_1=\eta\mu_r/2n, b_2=(\beta^2\tilde \lambda_{\text{min}}(I-W)/(\beta^2+1))-\eta^2 L_m$ and $\beta$ is a positive constant such that $\beta^2+2\beta=\frac{\mu_r}{2n\eta L_m}$.  
\end{lemma}}

\subsection{Main results}
{\color{black}The following lemmas are crucial for establishing a linear system of inequalities that  bound $\mathbb{E}_{\xi}[||\mathbf{X}^{k+1}-\mathbf{X}^\star||^2_{\text{F}}]$ and $\mathbb{E}_{\xi}[||\mathbf{X}^{k+1}-\mathbf{H}^{k+1}||^2_{\text{F}}]$ with respect to time $k$. 
 \begin{lemma}\label{mainl}
 	Given Assumption \ref{asp3}, \ref{c3} \ref{cvx}, \ref{asp1} and \ref{asp2}, when $\alpha \in(0,\frac{1}{r}]$, the following linear system of  component-wise inequalities holds
\begin{equation}
\begin{bmatrix}
	\mathbb{E}_{\xi}[||\mathbf{X}^{k+1}-\mathbf{X}^\star||^2_{\text{F}}\mid \mathcal{F}^k]\\

	\mathbb{E}_{\xi}[||\mathbf{X}^{k+1}-\mathbf{H}^{k+1}||^2_{\text{F}}\mid \mathcal{F}^k]\end{bmatrix}\preceq \mathbf{A}\begin{bmatrix}
	\mathbb{E}_{\xi}[||\mathbf{X}^{k}-\mathbf{X}^\star||^2_{\text{F}}\mid \mathcal{F}^k]\\

	\mathbb{E}_{\xi}[||\mathbf{X}^{k}-\mathbf{H}^{k}||^2_{\text{F}}\mid \mathcal{F}^k]\end{bmatrix},
	\end{equation}\label{ine}where the  elements of  matrix $\mathbf{A} = [a_{ij}]$ are given by
	
	\begin{equation}\label{a}
	\mathbf{A}=\begin{bmatrix}
	c_1(1+L_F^2\gamma^2-2\mu_F\gamma)&c_2\gamma^2\\
	c_3\gamma^2 L_F^2&c_x+c_4\gamma^2\\
	\end{bmatrix},
\end{equation}
with positive constants $c_i$'s and $c_x$  given in Appendix	\ref{ap1}.

 \end{lemma}
 \begin{proof}
   See Appendix	\ref{ap1}. 
\end{proof}}
\begin{remark}
In terms of the linear inequalities in Lemma \ref{mainl}, the optimzaition error $\mathbb{E}_{\xi}[||\mathbf{ X}^{k+1}-\mathbf{X}^\star||^2_{\text{F}}]$ and the compression error  $\mathbb{E}_{\xi}[||\mathbf{X}^{k+1}-\mathbf{H}^{k+1}||^2_{\text{F}}]$ all converge exponentially to $0$ if the spectral radius  $\rho(A)<1$.  The following lemma states a sufficient condition for ensuring $\rho(A)<1$. 
\end{remark}
\begin{lemma}\label{lm7}
	(Corollary 8.1.29 in \cite{horn2012matrix}) Let $\mathbf{M}\in \mathbb{R}^{n\times n}$ be a matrix with nonnegative elements and $\mathbf{v}\in \mathbb{R}^{n}$ be a vector with positive elements. If $\mathbf{Mv}\leq \theta \mathbf{v}$, there is $\rho(\mathbf{M})\leq \theta$. 
\end{lemma}

{\color{black}Before analyzing the convergence performance of C-DNES, we first give a theorem that proves the existence and uniqueness of NE  in game $\Gamma(n, \{J_i\}, \{\Omega_i\})$.
% Consider a convex, closed, and bounded set $S \subset \mathbb{R}^n$ and a function $\rho(\mathbf{x},\mathbf{y})$ defined by 
%     \begin{equation*}
%         \rho(\mathbf{x},\mathbf{y}) = \sum\limits_{i=1}^n J_i(x_1,\ldots,y_i,\ldots,x_n),
%     \end{equation*}
%     where $(\mathbf{x},\mathbf{y}) \in S\times S$. Under Assumption \ref{cvx}, it can be seen that $\rho(\mathbf{x},\mathbf{y})$ is continuous in $\mathbf{x}$ and $\mathbf{y}$ and is convex in $\mathbf{y}$ for every fixed $\mathbf{x}$.
%     Now we define a point-to-set mapping $\mathbf{x} \in S \rightarrow \Gamma\mathbf{x}\subset S$, given by 
% \begin{equation*}\Gamma\mathbf{x} =\{\mathbf{y} \mid  \rho(\mathbf{x},\mathbf{y})=\min_{\mathbf{z} \in S} \rho(\mathbf{x},\mathbf{z})\}.
% \end{equation*}
% From the continuity of $\rho(\mathbf{x},\mathbf{z})$ and the convexity in $\mathbf{z}$ of $\rho(\mathbf{x},\mathbf{z})$ for fixed $\mathbf{x}$,
%  it can be obtained that $\Gamma$ is an lower semicontinuous mapping that maps each point of the convex,
%  compact set $S$ into a closed convex subset of $S$ (Theorem 9.14 and Theorem 9.17 in \cite{Sundaram1996first}). Then by the Kakutani fixed point
%  theorem \cite{kakutani1941generalization}, there exists a point $\mathbf{x}^0 \in S$  such that $\mathbf{x}^0 \in \Gamma\mathbf{x}^0$,i.e.,
%  \begin{equation}\label{exi}
%     \rho(\mathbf{x}^0,\mathbf{x}^0)=\min_{ \mathbf{z} \in S} \rho(\mathbf{x}^0,\mathbf{z}).
%  \end{equation}
\begin{theorem}
Under Assumption \ref{cvx}, a Nash equilibrium $\mathbf{x}^\star$ exists in the game $\Gamma(n, {J_i}, {\Omega_i})$. Furthermore, under Assumption \ref{asp1}, this NE $\mathbf{x}^\star$ is unique.
\end{theorem}
\begin{proof}
  We first prove the existence of NE in  $\Gamma(n, {J_i}, {\Omega_i})$.  Consider a  function $\rho(\mathbf{x},\mathbf{y})$ defined by 
    \begin{equation*}
        \rho(\mathbf{x},\mathbf{y}) = \sum\limits_{i=1}^n J_i(x_1,\ldots,y_i,\ldots,x_n),
    \end{equation*}
    where $(\mathbf{x},\mathbf{y}) \in \mathbb{R}^n\times \mathbb{R}^n$. Under Assumption \ref{cvx}, it can be seen that $\rho(\mathbf{x},\mathbf{y})$ is continuous in $\mathbf{x}$ and $\mathbf{y}$ and is strongly convex in $\mathbf{y}$ for every fixed $\mathbf{x}$. Hence, $\rho(\mathbf{x},\mathbf{y})$ has a global minimum $\mathbf{x}^0 \in \mathbb{R}^n$ in $\mathbf{y}$, i.e.,
   
 \begin{equation}\label{exi}
    \rho(\mathbf{x}^0,\mathbf{x}^0)=\min_{ \mathbf{z} \in \mathbb{R}^n} \rho(\mathbf{x}^0,\mathbf{z}).
 \end{equation}

To show $\mathbf{x}^0$ is a NE $\mathbf{x}^\star$ in game $\Gamma(n, \{J_i\}, \{\Omega_i\})$, we will prove by contradiction. Assume for $i=l$ there exists a point $x_l'\ne x_l^0$ such that $\mathbf{x}'=(x_1^0,\ldots,x_l',\ldots, x_n^0) \in \mathbb{R}^n$ and $J_l(\mathbf{x}')<J_l(\mathbf{x}^0)$. Then we have $\rho(\mathbf{x}^0,\mathbf{x}')<\rho(\mathbf{x}^0,\mathbf{x}^0)$, which contradicts \eqref{exi}. Hence, the  minimum point  $\mathbf{x}^0$ is a NE   $\mathbf{x}^\star$  in game $\Gamma(n, \{J_i\}, \{\Omega_i\})$ satisfying \eqref{ne1}.

 Next, we show the uniqueness of NE under Assumption \ref{asp1}. If there exists a  NE    $\mathbf{x}^\star \in \mathbb{R}^n$ in  $\Gamma(n, \{J_i\}, \{\Omega_i\})$, we can obtain $\langle \mathbf{F}(\mathbf{x}^\star),\mathbf{x}-\mathbf{x}^\star\rangle \ge 0, \forall \mathbf{x}\in\mathbb{R}^n$ based on inequality \eqref{ne}. Assume there exists another NE $\mathbf{y}^\star$, we have  $\langle \mathbf{F}(\mathbf{x}^\star),\mathbf{y}^\star-\mathbf{x}^\star\rangle \ge 0$ and $\langle \mathbf{F}(\mathbf{y}^\star),\mathbf{x}^\star-\mathbf{y}^\star\rangle \ge 0$. Thus, $\langle \mathbf{F}(\mathbf{x}^\star)-\mathbf{F}(\mathbf{y}^\star),\mathbf{x}^\star-\mathbf{y}^\star\rangle \le 0$. Based on  Assumption \ref{asp1} that $\langle \mathbf{F(\mathbf{y}^\star)}-\mathbf{F(\mathbf{x}^\star)},\mathbf{\mathbf{y}^\star}-\mathbf{\mathbf{x}^\star}  \rangle \geq \mu_r ||\mathbf{\mathbf{y}^\star-\mathbf{x}^\star}||^2,$ we can derive that $\mathbf{x}^\star=\mathbf{y}^\star$. 
 
 In summary, there exists a unique NE  in game $\Gamma(n, \{J_i\}, \{\Omega_i\})$ under Assumption \ref{cvx} and \ref{asp1}.
\end{proof}}
The following theorem shows the convergence properties for the C-DNES algorithm in Algorithm \ref{alg1}.

%\begin{theorem}\label{th1}
%	Suppose Assumption \ref{asp3}, \ref{c3} \ref{cvx}, \ref{asp1} and \ref{asp2} hold, there exists a unique NE $bf{x}^\star$ in game $\Gamma(n, \{J_i\}, \{\Omega_i\})$. Moreover, when $\alpha \in(0,\frac{1}{r}]$, we have $$\mathbf{V}^{k+1}\le \mathbf{A}\mathbf{V}^{k},$$ 
%	
%	
%	where $\mathbf{V}^{k}=\begin{bmatrix}
%\mathbb{E}_{\xi}[||\mathbf{X}^{k+1}-\mathbf{X}^\star||^2_{\text{F}}]&\mathbb{E}_{\xi}[||\mathbf{X}^{k+1}-\mathbf{H}^{k+1}||^2_{\text{F}}]	
%\end{bmatrix}
%^\top$ and the  elements of transition matrix $\mathbf{A} = [a_{ij}]$ are  given by
%$$ \mathbf{A}=\begin{bmatrix}
%	\tau_1(1+L_F^2\gamma^2-2\mu_F\gamma)&c_2\\
%	\eta(c_3)&c_4\\
%	
%\end{bmatrix},$$
%with positive constants $c_i$'s and $c_x,t_x$ being given in Appendix	\ref{ap1}. In particular, if the consensus stepsize $\gamma=\mu_L/L_F^2$ and gradient stepsize $\eta$ satisfying
%	$$\eta \leq\min \left \{\frac{s(n+1)}{8\mu}\gamma, \frac{\mu \epsilon_2}{m_1 (n+1)}\gamma \right \},
%	$$  the optimization error $\mathbb{E}_{\xi}[||\mathbf{ X}^{k+1}-\mathbf{X}^\star||^2_{\text{F}}]$ and the compression error $mathbb{E}_{\xi}[||\mathbf{X}^{k+1}-\mathbf{H}^{k+1}||^2_{\text{F}}]$  both
% converge to $0$ at the linear rate $\mathcal{O}(\rho(\mathbf{A})^k)$, where $\rho(\mathbf{A})<1$. 
%\end{theorem} 
{\color{black}\begin{theorem}\label{th1}
	Under Assumption \ref{asp3}, \ref{c3} \ref{cvx}, \ref{asp1} and \ref{asp2} hold,  we take $\alpha \in(0,\frac{1}{r}]$,  the consensus step-size $\gamma=\mu_L/L_F^2$, and gradient step-size $\eta$ satisfying
	$$\eta\le \min\Big\{\frac{2n}{\mu_r}\sqrt{\frac{1-c_x}{m_1}},\frac{\tilde \lambda_{\text{min}}(I-W)m_2^2}{L_m(m_2^2+1)}\Big\},
	$$where $m_1,m_2$ are defined in \eqref{m1} and \eqref{m2}, respectively. Then the optimization error $\mathbb{E}_{\xi}[||\mathbf{ X}^{k+1}-\mathbf{X}^\star||^2_{\text{F}}]$ and the compression error $\mathbb{E}_{\xi}[||\mathbf{X}^{k+1}-\mathbf{H}^{k+1}||^2_{\text{F}}]$  both
 converge to $0$ at the linear rate $\mathcal{O}(\rho(\mathbf{A})^k)$ with $\rho(\mathbf{A})<1$, where  matrix $\mathbf{A}$ is defined in \eqref{a}. 
\end{theorem} 
 \begin{proof}Since $\gamma=\mu_F/L_F^2$,  we have $[\mathbf{A}]_{11}=c_1(1+L_F^2\gamma^2-2\mu_F\gamma)=(1-\mu_F^2/2L_F^2)<1$.  Recalling from Lemma \ref{lm7}, to ensure $\rho_{A}<1$, we can derive the range of $\eta$ and a positive vector $\mathbf{\epsilon}:=[\epsilon_1,\epsilon_2]^\top \in \mathbb{R}_{++}^2$, such that 
  \begin{equation}\label{pr1}
	\mathbf{A\epsilon}\le (1-\frac{\mu_F^2}{4L_F^2})\mathbf{\epsilon}
\end{equation}
holds. Below we determine conditions such that inequality \eqref{pr1} holds.

\textit{1) First inequality in \eqref{pr1}}: $\quad [\mathbf{A\epsilon}]_{1}\leq [(1-\frac{\mu_F^2}{4L_F^2})\mathbf{\epsilon}]_{1} $
\begin{equation}\label{ine1}
	(1-\frac{\mu_F^2}{2L_F^2})\epsilon_1+\frac{\mu_F^2 c_2}{L_F^4}\epsilon_2 \leq (1-\frac{\mu_F^2}{4L_F^2})\epsilon_1. 
\end{equation}
Inequality \eqref{ine1} holds if $\frac{4c_2}{L_F^2}\epsilon_2 \leq \epsilon_1$.

\textit{2) Second inequality in \eqref{pr1}}:  $\quad [\mathbf{A\epsilon}]_{2}\leq [(1-\frac{\mu_F^2}{4L_F^2})\mathbf{\epsilon}]_{2} $
 
\begin{equation}\label{ine3}
	\frac{c_3\mu_F^2}{L_F^2}\epsilon_1+(c_x+c_4\frac{\mu_F^2}{L_F^4})\epsilon_2\leq (1-\frac{\mu_F^2}{4L_F^2})\epsilon_2,
\end{equation}
which is equivalent to 
\begin{equation}
\frac{c_3\mu_F^2}{L_F^2}\epsilon_1+(\frac{\mu_F^2}{4L_F^2}+c_4\frac{\mu_F^2}{L_F^4})\epsilon_2\leq (1-c_x)\epsilon_2.
\end{equation}
Note that $L_F^2>||I-W||_{\text{F}}^2$ and denote 
\begin{equation}\label{m1}
	m_1=\frac{c_3\epsilon_1}{||I-W||_{\text{F}}^2}+\frac{\epsilon_2}{4||I-W||_{\text{F}}^2}+\frac{c_4\epsilon_2}{||I-W||_{\text{F}}^4}.
\end{equation} 
	Inequality \eqref{ine3} is verified with 
	\begin{equation}\label{muf}
		\mu_F\le \sqrt{\frac{1-c_x}{m_1}}
	\end{equation}

Since $\mu_F=\min\{b_1,b_2\}>0$, the inequality \eqref{muf} becomes
\begin{equation}
\begin{aligned}
		&b_1=\eta\mu_r/2n\le \sqrt{\frac{1-c_x}{m_1}}, \\
		&b_2=(\beta^2\tilde \lambda_{\text{min}}(I-W)/(\beta^2+1))-\eta^2 L_m>0.
\end{aligned}
\end{equation}
It suffices to have
\begin{equation}\label{m2}
	\eta\le \min\Big\{\frac{2n}{\mu_r}\sqrt{\frac{1-c_x}{m_1}},\frac{\tilde \lambda_{\text{min}}(I-W)m_2^2}{L_m(m_2^2+1)}\Big\},
\end{equation}
 where $m_2=-1+\sqrt{1+\frac{\mu_r^2}{4n^2L_m}\sqrt{\frac{m_1}{1-c_x}}}$.

To wrap up, if the positive constants $\epsilon_1, \epsilon_2$ and the step-size $\eta$ satisfy the following conditions, 
\begin{equation}
	\begin{split}
		&\epsilon_1\geq \frac{4c_2}{L_F^2}\epsilon_2,\quad \epsilon_2>0,\\
		&\eta\le \min\Big\{\frac{2n}{\mu_r}\sqrt{\frac{1-c_x}{m_1}},\frac{\tilde \lambda_{\text{min}}(I-W)m_2^2}{L_m(m_2^2+1)}\Big\},
	\end{split}
\end{equation}
the linear system of  element-wise inequalities in \eqref{pr1} can be shown and we can conclude that the optimization error $\mathbb{E}_{\xi}[||\mathbf{ X}^{k+1}-\mathbf{X}^\star||^2_{\text{F}}]$ and the compression error $\mathbb{E}_{\xi}[||\mathbf{X}^{k+1}-\mathbf{H}^{k+1}||^2_{\text{F}}]$  both
 converge to $0$ at the linear rate $\mathcal{O}(\rho(\mathbf{A})^k)$, where $\rho(\mathbf{A})<1$. 

\end{proof}
\begin{remark}
	It is worth noting that C-DNES can be equipped with different types of compressors  in different time iteration while existing quantized distributed NE seeking methods (\cite{chen2022distributed,ye2022distributed}) use a specific compressor for each time iteration. Furthermore, based on the mild assumption for compressors (Assumption \ref{c3}), we can even adopt multi-step compressions, such as the composition of quantization  and scarification $\mathcal{C}_q(\mathbf{\mathcal{C}_{top}(\mathbf{x})})$, to further reduce communication bits. In summary, C-DNES enjoys more flexibility in choosing compression methods.
	\end{remark}}
\begin{remark}
	All the above results can be adopted for games with different dimensions of the action sets. The scalar case is considered for the sake of notational simplicity.
\end{remark}
%%%%%%%%%%%%%%%%%%%%%%%%%%%%%%%%%%%%%%%%%%%%%%%%%%%%%%%%%%%%%%%%%%%%%%%%%%%%%%%%

\section{SIMULATIONS}
Consider a random generated directed communication network with $n=50$ agents. The weight matrix $W$ is defined as follows,

\begin{equation}
    [W]_{ij}=
   \begin{cases}
   1/(\max_{i \in V} \mathcal{|N|}_i),&\mbox{if $j \in \mathcal{N}_i$;}\\
   1-\sum_{j  \in \mathcal{N}_i}w_{ij},&\mbox{if $j=i$;}\\
   0,&\mbox{Otherwise.}
   \end{cases}
  \end{equation}

The connectivity control game defined in \cite{ye2020distributed} is considered, where the sensors in the network try to find a tradeoff between the local objective, e.g., source seeking and positioning and the global objective, e.g., maintaining connectivity with other sensors. The cost function of sensor $i$ is defined as follows,
\begin{equation}
\begin{aligned}
		J_i(\mathbf{x})=l_i^c(x_i)+l_i^g(\mathbf{x}),\\
\end{aligned}
	\end{equation}
	where $l_i^c(x_i)=x_i^\top r_{ii}x_i+x_i^\top r_i+b_i$, $l_i^g(\mathbf{x})=\sum\limits_{j\in S_i} c_{ij}||x_i-x_j||^2$,  $S_i\subseteq V$ and $x_i=[x_{i1},x_{i2}]^\top\in\mathbb{R}^2$ denotes the position of sensor $i$, $r_{ii},r_i, b_i, c_{ij}>0$ are constants.

In this simulation, the parameters are set as $r_{ii}=\begin{bmatrix}
	i&0\\0&i
\end{bmatrix}, r_i=\begin{bmatrix}i&i \end{bmatrix}^\top, b_i=i$ and $c_{ij}=1,    \forall i,j \in V$. Furthermore, there is $S_i=\{i+1\}$ for $i\in\{1,2,\ldots,49\}$ and $S_{50}=\{1\}$.
The  unique NE of the game is $x^\star_{ij}=-0.5$ for $i\in\{1,2,\ldots,50\}, j\in\{1,2\}$. Meanwhile,  $\mathbf{x}_{(i)}^0$ are randomly generated in $[0,1]^{100}$, $\mathbf{h}_{i}^0=\mathbf{0}$, the scaling paramater $\alpha$ and the consensus step-size $\gamma$ are set to $1$. 
    \begin{figure}[htp]
\centering
 \includegraphics[width=0.5\textwidth]{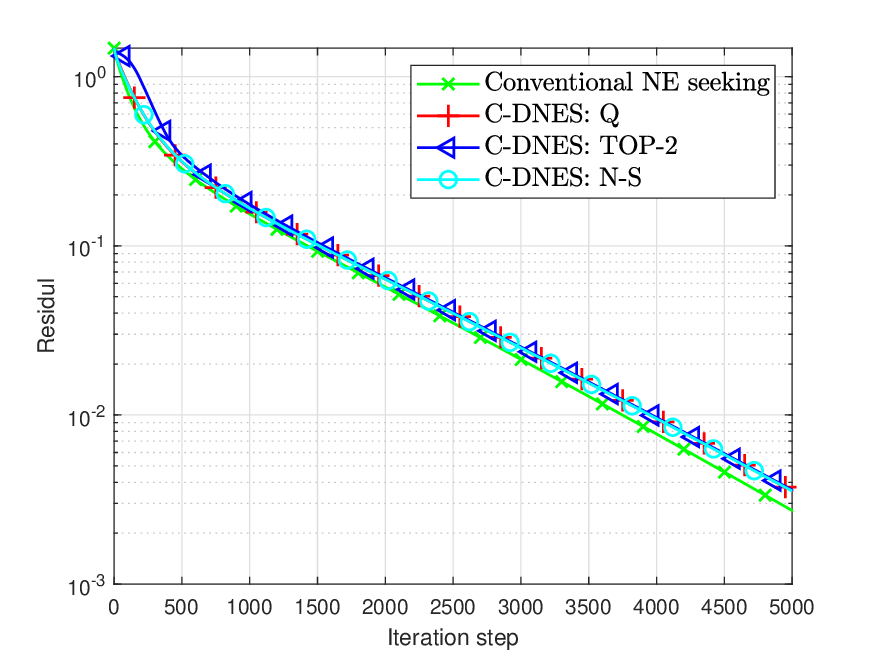}
  \caption{Residual ($||\mathbf{X}^k-\mathbf{X}^\star||_{\text{F}}$) v.s. number of iterations for the conventional distributed Nash equilibrium seeking algorithm \cite{tatarenko2020geometric} and the proposed C-DNES algorithm with different compressors.}
      \label{simu}
\end{figure}

  We tested three different compressors,  the stochastic quantization compressor  with $b=2$ and $q=\infty$ in \eqref{q1} (Q), the Top-k compressor with $k=2$ in \eqref{topk} (TOP-2) and the Norm-sign compressor with $q=\infty$ in \eqref{normsign} (N-S). By setting the step-size as $\eta=0.01$, Fig. \ref{simu} shows that C-DNES with different compressors all converge linearly to the NE  of the game and achieve   the same performance as the conventional distributed Nash equilibrium seeking algorithm without information compression \cite{tatarenko2020geometric}. Moreover, the effectiveness of different compressors is illustrated in Fig. \ref{nbit} and Fig. \ref{com_nbit}, which shows that Norm-sign compressor achieves the lowest communication cost among all algorithms. Furthermore, all compressed schemes  outperform 
 the conventional distributed NE seeking algorithm in terms of the communication burden. 
   \begin{figure}[htp]
\centering
 \includegraphics[width=0.5\textwidth]{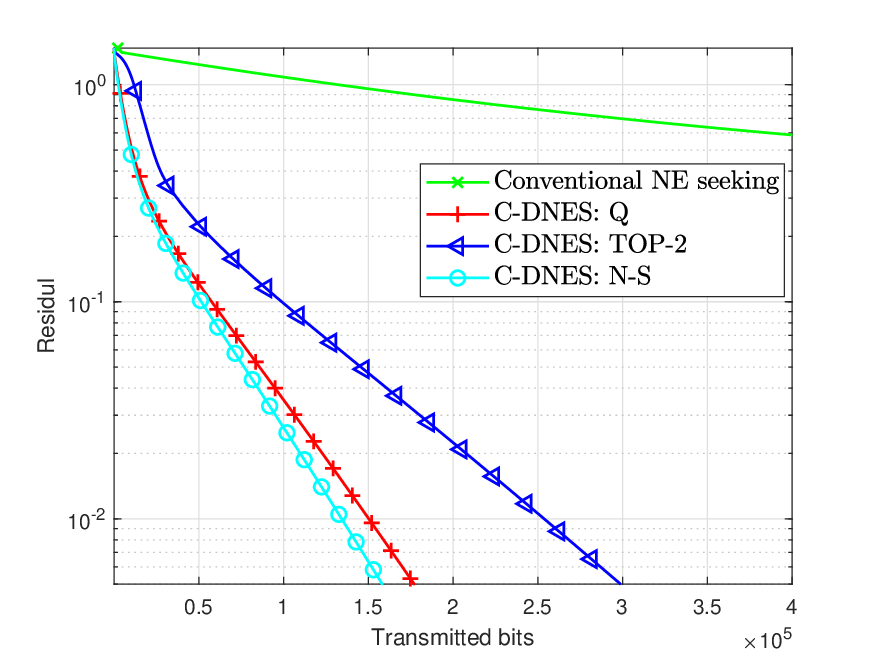}
  \caption{Residual ($||\mathbf{X}^k-\mathbf{X}^\star||_{\text{F}}$) v.s. transmitted bits for the conventional distributed Nash equilibrium seeking algorithm \cite{tatarenko2020geometric} and the proposed C-DNES algorithm with different compressors. }
      \label{nbit}
\end{figure}

   \begin{figure}[htp]
\centering
 \includegraphics[width=0.5\textwidth]{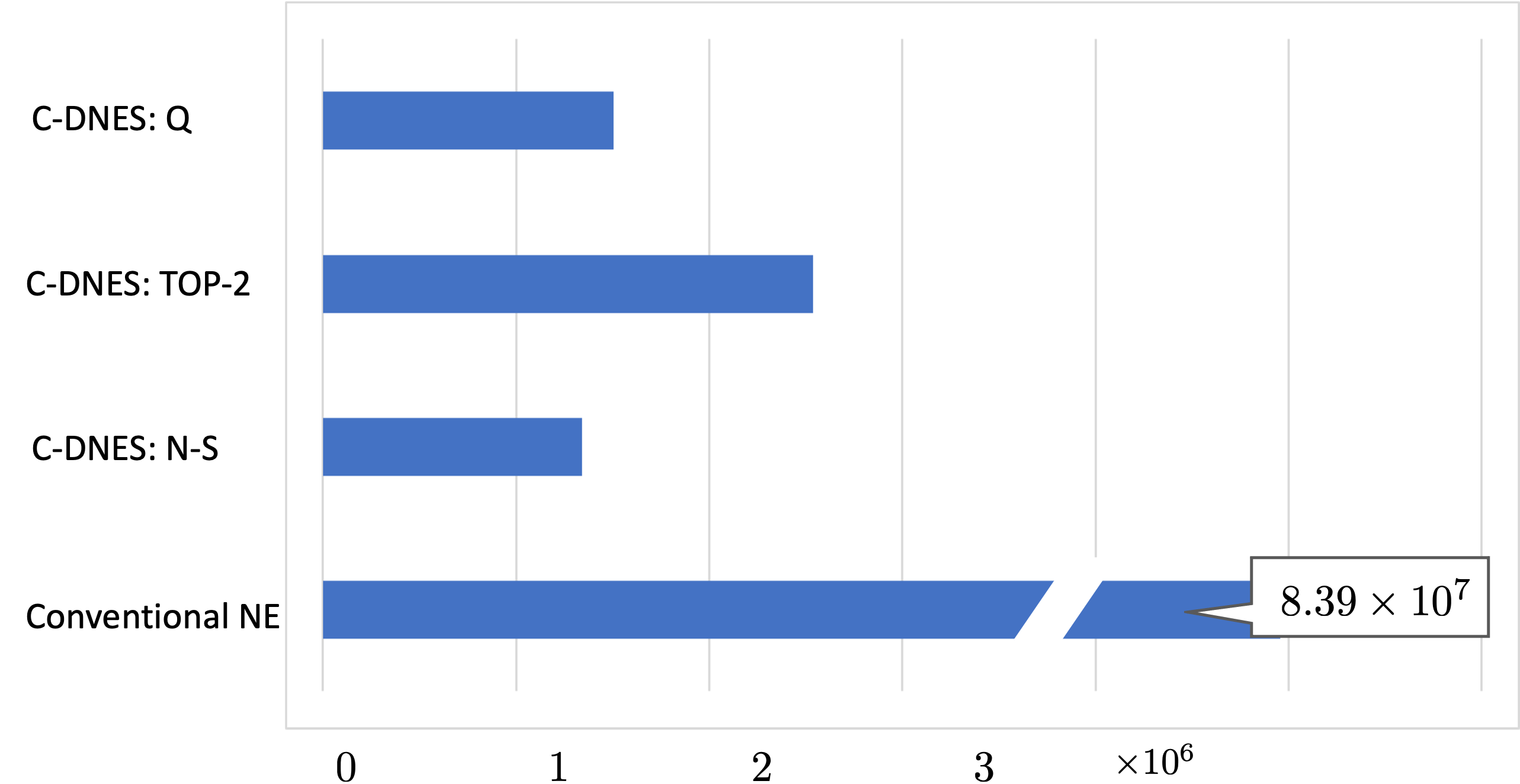}
  \caption{{\color{black}Total transmitted bits of different algorithms to obtain $\epsilon$-optimal NE with $\epsilon=10^{-2}$. }}
      \label{com_nbit}
\end{figure}

%%%%%%%%%%%%%%%%%%%%%%%%%%%%%%%%%%%%%%%%%%%%%%%%%%%%%%%%%%%%%%%%%%%%%%%%%%%%%%%%

\section {CONCLUSION AND FUTURE WORK}
{\color{black}In this paper, we study the problem of distributed Nash equilibrium seeking over a directed graph with communication information compression. Specifically, we propose a novel compressed distributed NE seeking approach (C-DNES) and prove its linear convergence. C-DNES not only  inherits the advantages of the conventional  distributed NE algorithm  for games with strongly monotone mappings, but also works with a general class of compression operators, e.g., unbiased and biased compressors.
Future works may focus on the extensions to networks over time-varying directed graphs. Distributed NE seeking with constrained action space will also be considered.  In addition, it is of interest to investigate the combination of acceleration techniques with C-DNES to further speed up the algorithm. }

\section{APPENDIX}{\color{black}
\subsection{Proof of Claim \ref{claim}}\label{cp}

From the property of vector norm, if $q\geq 2$, we have $||\mathbf{x}||_q^2\leq||\mathbf{x}||^2$ and $||\mathbf{x}||_q\geq \max\limits_i |x_i|$ for any vector $\mathbf{x}$.
\begin{equation}
	\begin{aligned}
		||\mathcal{C}( \mathbf{x})-\mathbf{x}||^2&=||||\mathbf{x}||_q sgn(\mathbf{x})-\mathbf{x}||^2\\
		&=\sum\limits_{i=1}^d(||\mathbf{x}||_q sgn(x_i)-x_i)^2\\
		&=\sum\limits_{i=1}^d(||\mathbf{x}||^2_q-2 ||\mathbf{x}||_qsgn(x_i)x_i+|x_i|^2)\\
		&\leq (d-2)||\mathbf{x}||^2_q+||\mathbf{x}||^2\\
		&\leq (d-1)||\mathbf{x}||^2.\\
	\end{aligned}
\end{equation}

\begin{equation}
	\begin{aligned}
		||\mathcal{C}^d( \mathbf{x})-\mathbf{x}||^2&=||\frac{||\mathbf{x}||_q sgn(\mathbf{x})}{d}-\mathbf{x}||^2\\
		&=\sum\limits_{i=1}^d(\frac{||\mathbf{x}||_q sgn(x_i)}{d}-x_i)^2\\
		&=\sum\limits_{i=1}^d(\frac{||\mathbf{x}||^2_q}{d^2}-\frac{2}{d} ||\mathbf{x}||_q|x_i|+|x_i|^2)\\
&\leq\sum\limits_{i=1}^d(\frac{||\mathbf{x}||^2_q}{d^2}-\frac{2}{d} |x_i|^2+|x_i|^2)\\
		&\leq \frac{1}{d}||\mathbf{x}||^2_q+||\mathbf{x}||^2-\frac{2}{d}||\mathbf{x}||^2\\
			&\leq(1-\frac{1}{d})||\mathbf{x}||^2.\\
	\end{aligned}
\end{equation}

\subsection{Proof of Lemma \ref{mainl}}\label{ap1}
We derive two inequalities in terms of NE-seeking error and compression error, respectively. 
 
 \textit{NE-seeking error:}    

Based on Proposition \ref{proposition} , we conclude that $\mathbf{X}^\star=\mathbf{X}^\star-\gamma\mathbf{F}_a(\mathbf{X}^\star), \forall \gamma>0$. 

Thus,   we can obtain 
\begin{equation}\label{eqxstar1}
	\begin{aligned}
		&||\mathbf{X}^{k+1}-\mathbf{X}^\star||_{\text{F}}^2\\
		&\le||\mathbf{X}^{k}-\gamma \mathbf{F}_a(\mathbf{X}^{k})+\gamma(I-W)\mathbf{ E}^{k}-\mathbf{X}^\star+\gamma\mathbf{F}_a(\mathbf{X}^\star)||_{\text{F}}^2\\
		&\le \tau_1||\mathbf{X}^{k}-\gamma \mathbf{F}_a(\mathbf{X}^{k})-\mathbf{X}^\star+\gamma\mathbf{F}_a(\mathbf{X}^\star)||_{\text{F}}^2 \\
		&+\frac{\tau_1}{\tau_1-1}\gamma^2||I-W||_{\text{F}}^2||\mathbf{ E}^{k}||_{\text{F}}^2,
	\end{aligned}
\end{equation} 
where the second inequality comes from Lemma \ref{lmeq} with $\tau_1>1$. 

Next, we bound
\begin{equation}\label{eqxstar}
	\begin{aligned}
		&||\mathbf{X}^{k}-\gamma \mathbf{F}_a(\mathbf{X}^{k})-\mathbf{X}^\star+\gamma\mathbf{F}_a(\mathbf{X}^\star)||_{\text{F}}^2\\
		&=||\mathbf{X}^{k}-\mathbf{X}^\star||_{\text{F}}^2+\gamma^2|| \mathbf{F}_a(\mathbf{X}^{k})-\mathbf{F}_a(\mathbf{X}^\star)||_{\text{F}}^2\\
	&-2\gamma\langle \mathbf{F}_a(\mathbf{X}^{k})-\mathbf{F}_a(\mathbf{X}^\star),\mathbf{X}^{k}-\mathbf{X}^\star\rangle\\
	&\le (1+\gamma^2L_F^2-2\gamma\mu_F)||\mathbf{X}^{k}-\mathbf{X}^\star||_{\text{F}}^2,
	\end{aligned}
\end{equation}
where the last inequality is based on Lemma \ref{lml} and Lemma \ref{lmmu}. 

From \eqref{eqxstar},  we have $\min\{1+\gamma^2L_F^2-2\gamma\mu_F\}=(1-\mu_F^2/L_F^2)>0$. Combining \eqref{eqxstar1} and \eqref{eqxstar}, we can have the following inequality by taking $\tau_1=\frac{2L_F^2-\mu_F^2}{2L_F^2-2\mu_F^2}$.

\begin{equation}
	\begin{aligned}
		||\mathbf{X}^{k+1}-\mathbf{X}^\star||_{\text{F}}^2&\le c_1(1+\gamma^2L_F^2-2\gamma\mu_F)||\mathbf{X}^{k}-\mathbf{X}^\star||_{\text{F}}^2 \\
		&+\frac{c_1\gamma^2||I-W||_{\text{F}}^2}{c_1-1}||\mathbf{ E}^{k}||_{\text{F}}^2,
	\end{aligned}
	\end{equation}
	where $c_1=\frac{2L_F^2-\mu_F^2}{2L_F^2-2\mu_F^2}$.
	
   Then, from Lemma \ref{lmm}, we can obtain
   \begin{equation}\label{nee}
	\begin{aligned}
		&\mathbb{E}_{\xi}[||\mathbf{X}^{k+1}-\mathbf{X}^\star||_{\text{F}}^2\mid \mathcal{F}^k]\\
		&\le c_1(1+\gamma^2L_F^2-2\gamma\mu_F)\mathbb{E}_{\xi}[||\mathbf{X}^{k}-\mathbf{X}^\star||_{\text{F}}^2\mid \mathcal{F}^k] \\
		&+c_2\gamma^2\mathbb{E}_{\xi}[||\mathbf{X}^{k}- \mathbf{H}^{k}||^2\mid \mathcal{F}^k],
	\end{aligned}
	\end{equation}
	where $c_2=\frac{c_1||I-W||_{\text{F}}^2C}{c_1-1}$. 
   
    \textit{Compression error of the decision variable:} 
    
    Denote $\mathcal{C}_r^k(\mathbf{X}^{k})=\mathcal{C}^k(\mathbf{X}^{k})/r$,  according to \eqref{ald}, for $0<\alpha\leq \frac{1}{r}$, we have 
    \begin{equation}\label{cpe1}
    \begin{split}
    	&    	||\mathbf{X}^{k+1}- \mathbf{H}^{k+1}||^2\\
    	=&||\mathbf{X}^{k+1}-\mathbf{X}^{k}+\mathbf{X}^{k}- \mathbf{H}^{k}-\alpha r\frac{\mathbf{Q}^{k}}{r}||^2\\
    	=&||\mathbf{X}^{k+1}-\mathbf{X}^{k}+\alpha r(\mathbf{X}^{k}- \mathbf{H}^{k}-\mathcal{C}_r^k(\mathbf{X}^{k}- \mathbf{H}^{k}))\\
    	&+(1-\alpha r)(\mathbf{X}^{k}- \mathbf{H}^{k})||^2\\
    	\leq &\tau_2||\alpha r(\mathbf{X}^{k}- \mathbf{H}^{k}-\mathcal{C}_r^k(\mathbf{X}^{k}- \mathbf{H}^{k}))\\
    	&+(1-\alpha r)(\mathbf{X}^{k}- \mathbf{H}^{k})||^2+\frac{\tau_2}{\tau_2-1}||\mathbf{X}^{k+1}-\mathbf{X}^{k}||^2\\
    	\leq &\tau_2\Big [\alpha r||\mathbf{X}^{k}- \mathbf{H}^{k}-\mathcal{C}_r^k(\mathbf{X}^{k}- \mathbf{H}^{k})||^2\\
    	&+(1-\alpha r)||\mathbf{X}^{k}- \mathbf{H}^{k}||^2\Big ]+\frac{\tau_2}{\tau_2-1}||\mathbf{X}^{k+1}-\mathbf{X}^{k}||^2,\\
    \end{split}
    \end{equation}
    where in the first inequality we use the result of Lemma \ref{lmeq} with $\tau_2>1$. 
    
    Taking conditional expectation on both sides of \eqref{cpe1}, we obtain
\begin{equation}\label{cpe2}
	\begin{split}
		    	&\mathbb{E}_{\xi}[||\mathbf{X}^{k+1}- \mathbf{H}^{k+1}||^2\mid \mathcal{F}^k]\\
    	&\leq \tau_2[\alpha r (1-\delta)+(1-\alpha r)]\mathbb{E}_{\xi}[||\mathbf{X}^{k}- \mathbf{H}^{k}||^2\mid \mathcal{F}^k]\\
    	&+\frac{\tau_2}{\tau_2-1}\mathbb{E}_{\xi}[||\mathbf{X}^{k+1}-\mathbf{X}^{k}||^2\mid \mathcal{F}^k],\\ 
	\end{split}
\end{equation}
where the inequality holds based on Assumption \ref{c3}. 

Moreover, we have
\begin{equation}\label{cpe3}
\begin{split}
		&\mathbb{E}_{\xi}[||\mathbf{X}^{k+1}-\mathbf{X}^{k}||^2\mid \mathcal{F}^k]\\
		&=\mathbb{E}_{\xi}[||\gamma(I-W)(\mathbf{X}^{k}-\mathbf{\hat X}^{k})-\gamma\mathbf{F}_a(\mathbf{X}^k)+\gamma\mathbf{F}_a(\mathbf{X}^\star)\mid \mathcal{F}^k]\\
		&\leq 2\gamma^2C||(I-W)||^2 \mathbb{E}_{\xi}[||\mathbf{X}^{k}-\mathbf{H}^{k}||^2\mid \mathcal{F}^k]\\
		&+2\gamma^2 L_F^2\mathbb{E}_{\xi}[||\mathbf{X}^{k}- \mathbf{X}^{\star}||^2\mid \mathcal{F}^k] \\
		\end{split}
\end{equation}
Bringing \eqref{cpe3} into \eqref{cpe2} and denoting $c_3=\frac{2\tau_2 }{\tau_2-1}>1, c_x=\tau_2[\alpha r (1-\delta)+(1-\alpha r)]=\tau_2(1-\alpha r\delta)<1$, we have
\begin{equation}\label{cpe4}
	\begin{split}
		    	&\mathbb{E}_{\xi}[||\mathbf{X}^{k+1}- \mathbf{H}^{k+1}||^2_{\text{F}}\mid \mathcal{F}^k]\\
		    	&\leq c_3\gamma^2 L_F^2\mathbb{E}_{\xi}[||\mathbf{X}^{k}-\mathbf{X}^{\star}||_{\text{F}}^2\mid \mathcal{F}^k]\\
    	&+ (c_x+c_4\gamma^2)\mathbb{E}_{\xi}[||\mathbf{X}^{k}- \mathbf{H}^{k}||^2\mid \mathcal{F}^k],\\
	\end{split}
\end{equation}
where $c_4=c_3C||(I-W)||^2_{\text{F}}$. 

Combing the above equalities \eqref{nee} and \eqref{cpe4} yields the desired result.

}
\end{document}